\documentstyle[prl,aps]{revtex}
\begin{document}

\draft
\title{Adiabatically changing the phase-space density 
 of a trapped Bose gas}
\author{P.W.H. Pinkse, A. Mosk,  M. Weidem\"{u}ller $^*$, 
 M.W. Reynolds, T.W. Hijmans, and J.T.M. Walraven
\thanks{Present address: FOM Institute for Atomic 
and Molecular Physics (AMOLF), Kruislaan 407, 
1098 SJ Amsterdam, The Netherlands}
}
\address{Van der Waals -- Zeeman Institute,
University of Amsterdam,\\
Valckenierstraat 65/67, 1018 XE Amsterdam, The Netherlands}
\date{\today}

\maketitle

\begin{abstract}
We show that 
the degeneracy parameter 
of a trapped Bose gas can be changed adiabatically 
in a reversible way, 
both in the Boltzmann regime
and in the degenerate Bose regime.
We have performed measurements on
spin-polarized atomic hydrogen 
in the Boltzmann regime demonstrating reversible
changes of the degeneracy parameter (phase-space density)
by more than a factor of two. This result is in perfect 
agreement with theory. 
By extending our theoretical analysis to the quantum degenerate regime we predict
that, starting close enough to the Bose-Einstein 
phase transition, one can cross the transition by 
an adiabatic change of the trap shape.
\end{abstract}

\narrowtext
\vspace{1cm}

The observation of
Bose-Einstein condensation (BEC) in
magnetically trapped
atomic vapors of rubidium \cite{Anderson}, sodium \cite{Davis},
and lithium \cite{Bradley}
has opened a new field of study at the intersection of atomic
and condensed matter physics.
Presently, condensates are produced routinely and 
detailed studies of condensate properties,
such as collective excitations \cite{Jin96} and interaction
of two condensates \cite{Ketterle96},  are being made.
The BEC phase transition itself is especially intriguing.
Open questions include the kinetics
of condensate formation and the effect of interatomic
interactions and finite number of particles.
Thus far, measurements have relied upon evaporative cooling
to prepare the sample at the desired density below the
critical temperature.
Evaporative cooling, however, is inherently
irreversible since it is based on the loss of hot particles 
from the trap \cite{evaporationrefs,Luiten96}.  To tackle the above
questions it would be extremely valuable to vary the
degeneracy of the trapped gas adiabatically in a reversible manner,
with a fixed number of particles,
especially since non-destructive detection methods
have become available \cite{darkground}.

The possibility of increasing phase-space density and reaching BEC by
changing the trapping potential was investigated by Ketterle
and Pritchard \cite{KetterleandPritchard}.
For a collisionless gas they showed that it is impossible to
influence phase-space density by manipulating the trapping potential.
An example is cooling by adiabatic expansion 
in a harmonic trap:
one does not lose any atoms, but
one does not get closer to BEC either.

In this paper we show that this ``no pain, no gain"
principle is not true for a collisional gas.
We experimentally demonstrate that the 
degeneracy parameter 
of a trapped gas can be changed adiabatically (without exchange of heat)
and reversibly, 
without sacrificing atoms, by changing the shape of the trap
slowly compared to the internal equilibration time.
Our experiments are done in the Boltzmann regime where
the degeneracy parameter $n \Lambda^3$
coincides with phase-space density.
Here $n$ denotes the density of the gas 
at the minimum of the potential
and $\Lambda\equiv (2\pi \hbar^2 /m T)^{1/2}$ 
is the thermal de Broglie wavelength at temperature $T$
(with $m$ the
atomic mass and Boltzmann's constant $k_{\rm B}=1$).
We observed a change of 
$n \Lambda^3$ by a factor of two,
which agrees well with 
a quantitative prediction based on 
statistical thermodynamics applied to a trapped gas. 
We point out that, given suitable starting conditions,
this approach is also suited to cross the BEC phaseline. 
This follows from analytical expressions 
for heat capacity, entropy and condensate fraction
of a Bose gas as a function of the trap shape.

Our experiment is performed with atomic hydrogen
in the cryogenic Ioffe trap described by Van Roijen 
{\it et al} \cite{Trap}. 
To determine quantities like temperature and density 
we measure the Lyman-$\alpha$ absorption spectrum of the gas
and fit
calculated spectra to the experimental one
(see \cite{Luiten93} for details). 
It takes a measuring time of 40 seconds to reach
a $10\%$ level of accuracy under present conditions.
To minimize heating by photon recoil,
we used our Lyman-$\alpha$ source at low intensity
(typically $10^{6}$ photons at the sample 
per pulse, 20 pulses per second).
In order to assure sufficient signal to noise ratio,
we replaced the photodiode used in previous 
experiments \cite{Luiten93}
by a photomultiplier. 
This method offers a non-destructive way 
to  follow the evolution of
a trapped gas {\em in situ} under changing conditions.
 
The character of the Ioffe trap can be controlled via the
magnetic field at the trap center, $B_0$.  
For $\mu_B B_0\gg T$ (with $\mu_B$ the Bohr magneton), 
the trap is essentially harmonic.
For $\mu_B B_0\ll T$, the trap
is close to linear in the radial direction and nearly harmonic axially.
We selected two magnetic field configurations, {\bf A} and {\bf B}, 
shown in Fig. \ref{geometries},
that maximize the change in phase-space density 
within the constraints of our apparatus.

After loading the trap from a cryogenic dissociator 
the sample is evaporatively cooled
by ramping down the magnetic barrier at one 
of the longitudinal ends of the trap 
(at $z=5$ cm, see Fig. \ref{geometries}). 
Evaporation is then
stopped by raising this barrier. 
Hereafter we let the sample decay and equilibrate 
for about $1000$~s
to prepare a thermal sample that is better than
95\% doubly spin polarized with typical
densities around  $10^{11}$ atoms/cm$^3$ \cite{Trap}.
This density was selected to assure negligible sample 
loss during measurement of the spectrum.
We gradually change from trap {\bf A} to trap {\bf B} 
or vice versa in 60~s 
which is much slower than the average elastic collision 
time of 5~s at this density.
After each change of trap a spectrum is measured to determine
the number of atoms and the degeneracy parameter. 

Changing the trap configuration from {\bf A} to {\bf B},
the temperature increased reversibly from 
$T_{A} \approx 46$mK to $T_{B} \approx60 $mK.
Both $T_{A}$ and $T_{B}$ showed 
an upward drift of $\sim 10$mK to the
final values mentioned above reached after $5000$~s.
In Fig.~\ref{results} we plot the degeneracy parameter
$n \Lambda^3$ versus time  for 
a sample cycled between trap {\bf A} and trap {\bf B}.
It can be seen that 
the phase-space density
differs by a constant factor $2.05 \pm 0.13$ between 
trap {\bf A} and trap {\bf B}.
The number of trapped atoms $N$ versus time is plotted
in the lower graph of  Fig.~\ref{results}.
Although $T$ and $n$ differ considerably between the two traps,
the measured value for $N$ is 
seen to be conserved within experimental error.
The slow decay of $N$ was found to be described by 
$N(t)=N(0)/[1+N(0) G t]$ as one would expect for a 
second-order decay process.
The rate $G=(1.34\pm0.05) \times 10^{-14}$ per atom per second
is consistent with dipolar relaxation
\cite{Trap,decaytheory}. 

Our results can be understood within a dilute gas model
for $N$ atoms at temperature $T$
trapped in a deep external potential, 
so that evaporation is negligible.
Collisions keep the gas in internal thermal equilibrium. 
The number of particles in the gas is
sufficient to enable a thermodynamic description.
Since we change the trap potential slowly compared to
the thermalization time, 
thermodynamic processes proceed reversibly.
Since there is no exchange
of heat or particles with the environment, 
the thermodynamic entropy $S$ of the gas is constant.
In our experiment, the influence
of quantum statistics is small and
interactions between atoms do not influence thermodynamics 
as the mean-field interaction energy
is much smaller than $T$.
The degeneracy parameter can be expressed in terms of 
the single-particle partition function $Z_1$ 
and the total number of particles $N$ as
$n\Lambda^3=N/Z_1$ \cite{evaporationrefs}.

The internal energy $E$ can be
calculated from the partition function to give \cite{Luiten96}
\begin{equation}
\label{eboltz}
	E=\left( \frac{3}{2} +\gamma \right) N T,
\end{equation}
where $\gamma=(T/V_e) \partial V_e/\partial T$,
with $V_e \equiv N/n=Z_1 \Lambda^3$ the effective volume.
Here $\gamma T$ is the average potential energy per particle.
For many trapping potentials 
$\gamma$ is a constant independent of $T$
and $V_e$ scales like $T^\gamma$.
For example, for a box $\gamma=0$, 
for a harmonic trap $\gamma=3/2$,
and for a spherical quadrupole trap $\gamma=3$.
The canonical partition function $Z_1^N/N !$
can be written as the exponential of the
Helmholtz free energy $E-TS$ (see, e.g., \cite{Reif})
to arrive at the following expression for the
degeneracy parameter:
\begin{equation}
\label{zisexp}
	n\Lambda^3=\exp
	\left(\frac{5}{2} + \gamma - \frac{S}{N} \right).
\end{equation}
Since in an adiabatic process $S$ and $N$ are constant,
the phase-space density does not change unless $\gamma$ changes.
It immediately follows  that many of 
the most obvious ways of changing the
trapping potential,
like isotropic or anisotropic scaling of a harmonic trap,
do {\em not} influence phase-space density. 
If we change $\gamma$ adiabatically
(e.g., experimentally, by changing
$B_0$ in a Ioffe trap), $n\Lambda^3$ will
change as $e^{\gamma}$.
In an extreme case, by slowly changing the trap shape from
square well to spherical quadrupole the
phase-space density in the center of the trap can be increased by as much as 
a factor $e^3 \approx 20$.

For the two traps used in our experiment 
the value of $\gamma$ is weakly temperature dependent.
For the measured temperatures we find 
$\gamma_{\rm A}\approx1.79$ for trap {\bf A} and 
$\gamma_{\rm B}\approx2.53$ for trap {\bf B}, 
the difference $\Delta \gamma=\gamma_{\rm B}-\gamma_{\rm A}$
after changing trap shape 
always being $0.74\pm0.01$.
This implies a 
change in phase-space density by a factor 
$\exp[\Delta \gamma]=2.10\pm0.02$
in perfect agreement with the measured value
of $2.05\pm0.13$.

We emphasize that it should be possible to achieve
considerable changes in the degeneracy parameter
also around the BEC transition.
For this purpose we
extend our theoretical consideration to the case of
a noninteracting Bose gas
in the degenerate regime.
This is a good approximation also for a weakly interacting Bose gas
as long as the gas parameter
$na^3\ll1$, where $a$ is the scattering length.
Below the critical temperature we have the additional condition
that $n_0 \tilde{U}\ll T$, 
with $\tilde{U}=4 \pi \hbar^2 a/m$ the scattering strength 
and $n_0$ is the condensate density.
Expressions obtained for the entropy of an ideal Bose gas 
are still good approximations
around and above the critical temperature of a non-ideal
Bose gas.
Interactions will change the shape of the condensate, 
but its entropy will always be zero. 
Because the influence of the interactions on the entropy of 
the above-condensate particles is negligible 
in the binary collision regime
the interactions only become important when the condensate fraction
becomes so large that the mean field energy of the condensate
changes the effective potential for the above-condensate particles.
Gases used
in current investigations, such as Rb, Li, Na and H
are sufficiently close to this ideal gas limit that
thermodynamics are essentially unaffected by interactions,
unless the condensate fraction becomes appreciable.

In the degenerate regime
we continue to assume quasiclassical
motion of the atoms. For clarity we restrict ourselves to 
the case of a power-law potential, 
although our expressions can easily be generalized to 
include the Ioffe trap.
A power-law trap is characterized by a density of states of the type
$\rho(\epsilon)=A \epsilon^{{1}/{2} + \delta}$ 
(See \cite{Powerlaw} and \cite{Bagnato} for details).
The scaling parameter $A$ 
determines the size of the trap. E.g. for a harmonic trap
$A=\frac{1}{2} (\hbar \omega)^{-3}$, 
where $\omega$ is the trap frequency.
The parameter $\delta$ governs the shape of the trap.
It can be shown that in the Boltzmann regime
$\gamma=\delta$ independent of $T$ for all
power-law traps. 

The internal energy 
of a Bose gas in a power-law trap
above and below the critical temperature $T_c$\cite{explosion}
is given by
\begin{mathletters}
\begin{eqnarray}
	E&=&NT
\left(
 \frac{3}{2}+\delta
\right)
	\frac{g_{\frac{5}{2}+\delta}(z)}{g_{\frac{3}{2}+\delta}(z)}\ \
	; T \geq T_c, \label{Ebose1} \\
	E&=&NT
\left(
\frac{3}{2}+\delta
\right)
	\frac{g_{\frac{5}{2}+\delta}(1)}{g_{\frac{3}{2}+\delta}(1)}
	\left( \frac{T}{T_c} \right)^{\frac{3}{2}+\delta}\ \
	; T \leq T_c, \label{Ebose2}
\end{eqnarray}
\end{mathletters}
where the Bose-Einstein integrals 
are expressed in polylogarithms 
$g_\alpha(x)
=\sum_{l=1}^\infty x ^l l^{-\alpha}$.
The fugacity 
$z=\exp(\mu/T)$ with $\mu$ the chemical
potential \cite{Huang}. 
Note that as we have $T_c^{3/2 +\delta}\propto N$,
Eq.~(\ref{Ebose2}) is in fact independent of the number of atoms. 
Above $T_c$ the fugacity is given implicitly by
\begin{equation}
	N=AT^{\frac{3}{2}+\delta} \Gamma(\frac{3}{2}+\delta)
		g_{\frac{3}{2}+\delta}(z),
\end{equation}
with $\Gamma(x)$ the Euler gamma function.
Below $T_c$, $z=1$ and the number of atoms
in the condensate, $N_0$, is given by 
$N_0/N = 1-(T/T_c)^{3/2+\delta}$.
In the high temperature (Boltzmann) limit the 
parameter $n \Lambda^3$ introduced earlier
reduces to the fugacity $z$.
The specific heat at constant particle number and
constant trap potential 
can now be found by
taking the derivative of $E$ with respect to temperature. We obtain
\begin{mathletters}
\begin{eqnarray}
	C&=& N 
\left(
  \frac{3}{2}+\delta
\right)
		\left(
		f_{\frac{5}{2}+\delta}(z)
		-f_{\frac{3}{2}+\delta}(z)
		\right)
		; \ \ T > T_c, \label{heatcap1} \\
	C&=& N 
\left(
  \frac{3}{2}+\delta
\right)
		f_{\frac{5}{2}+\delta}(1)
		\left( \frac{T}{T_c} \right)^{3/2+\delta}
			; \ \ T < T_c, \label{heatcap2}
\end{eqnarray}
\end{mathletters}
where we have introduced
$
	f_{\kappa}(z) \equiv
	\kappa g_{\kappa}(z)/g_{\kappa-1}(z)
$. 
For $\delta=0$, $\frac{3}{2}$ and $3$ we find
$f_{5/2+\delta}(1)=$ 1.284, 3.602 and 5.346, respectively.
In Eq.~(\ref{heatcap1}) the second term gives rise to a discontinuity
in the heat capacity at $T_c$ as already found by 
Bagnato et al. \cite{Bagnato},
which appears only for 
$\delta>1/2$ because for $\delta<1/2$ the function
$g_{1/2+\delta}$ diverges, and hence $f_{3/2+\delta}$ tends to zero,
as $T \rightarrow T_c$.
This jump in the heat capacity was recently observed by the JILA
group \cite{Wieman}.
The entropy $S$ obeys $dS/dT=C/T$, 
and hence has a kink at $T_c$ for $\delta>1/2$.
We find that
\begin{mathletters}
\label{entropy}
\begin{eqnarray}
	S &=& N 
\left(
f_{\frac{5}{2}+\delta}(z) 
		-\ln(z) 
\right)
		; \ \ T \geq T_c, \label{entropy1} \\
	S &=& N f_{\frac{5}{2}+\delta}(1) 
		\left( \frac{T}{T_c} \right)^{\frac{3}{2}+\delta} 
		; \ \ T \leq T_c. \label{entropy2} 
\end{eqnarray}
\end{mathletters}
For the homogeneous case ($\delta=0$) 
these expressions can be found in standard
statistical mechanics textbooks (e.g.\cite{Huang}).
The entropy as a function of $T/T_c$ is plotted 
in Fig. \ref{entropyfig}. 

It is noteworthy that the ideal Bose gas in a power-law trap 
in three space dimensions is
isomorphic to the uniform Bose gas in $2\delta+3$ dimensions.
There is no heat capacity jump when the effective dimension
is less than four. 
Remarkably, the processes which change the degeneracy parameter
are just those which correspond to adiabatic changes of dimensionality!

If one moves along an isentropic line
in Fig. \ref{entropyfig} by increasing $\delta$, 
one can take a dilute gas close to BEC through the transition.
For example, by varying $\delta$ from $0$ to $3$ it is possible
to Bose condense a gas that had initially 
a temperature $12$ times higher than the critical temperature
for that trap and given $N$. 
Eqs. (\ref{entropy})
allow one to calculate the fraction of condensate
particles $N_0/N$. 
Starting with a Bose gas in an ideal harmonic potential at 
$T=T_c$, and reversibly changing the trapping potential to an ideal
spherical-quadrupole, we would arrive below $T_c$ with
a condensate fraction of $0.33$.
For the Ioffe trap, starting at 
$T=T_c$ at the limit of high $B_0$, 
we expect a condensate
fraction of 0.25 in the trap at the limit of low $B_0$.
Of course in these case $T/T_c$ may already be so low 
that the influence of interactions can no longer be neglected.

Our method allows one to control the degeneracy of a Bose gas 
and gently pull it across $T_c$ and back.
Cycling times are only 
limited by the elastic collision rate which for the alkali atoms 
has been demonstrated 
to be much faster than for hydrogen.
Combined with non destructive diagnostics \cite{darkground}
this yields the unique possibility to study both
condensate formation and destruction and
to establish the presence or absence of asymmetries and hysteresis.

This work is part of a research program of the 
Stichting voor Fundamenteel Onderzoek der Materie (FOM),
which is a subsidiary of the Nederlandse Organisatie voor
Wetenschappelijk Onderzoek (NWO). M.W.\ acknowledges a TMR 
grant by the European Commission. The research of M.W.R.\
is supported by the Royal Netherlands Academy of Arts 
and Sciences (KNAW).

\begin{figure}
	\vspace{1cm}
	\caption{
	 Magnetic field profile of trap {\bf A} and {\bf B}
	 radially cut through the field minimum (left) and
	 cut along the principal axis of the trap (right).
         The horizontal lines indicate the measured thermal energy $T$.
         Trap {\bf A} has a depth of 0.72 K and a field minimum 
         $B_0$ of 226 mT,
	 Trap {\bf B} has a depth of 0.85 K and a field minimum $B_0$ of 12 mT.
	 }
	\label{geometries}
\end{figure}

\begin{figure}
	\vspace{1cm}
	\caption{Experimentally determined phase-space densities 
	 of trapped atomic hydrogen (upper graph), 
	 and number of atoms $N$ in the trap (lower graph)
         as a function of time,
	 while the trapping potential was 
	 alternated between trap {\bf A} and {\bf B}. The curves are 
	 fits to a second order decay, 
	 with the constant ratio $2.05 \pm 0.13$
	 of phase-space density
	 between the two traps.
	 }
	\label{results}
\end{figure}

\begin{figure}
	\vspace{1cm}
	\caption{
	 Entropy of an ideal Bose gas in power-law traps as a function of
	 $T/T_c$ (right) and, below $T_c$, 
	 also as a function of the above condensate fraction (left).
         }
	\label{entropyfig}
\end{figure}


\begin{references}
\bibitem{Powerlaw}
  This includes the important examples of
  square ($\delta=0$),
  spherical-quadrupole ($\delta=3$) 
  and harmonic traps ($\delta=3/2$),
  the low and high $B_0$ limits of the Ioffe trap \cite{Luiten96}
  and all traps with potentials that
  can be written as $U(x,y,z)=
  \mbox{$a\mid\!\!x\!\!\mid^{1/\delta_1}\!$}+
  \mbox{$b\mid\!\!y\!\!\mid^{1/\delta_2}\!$}+
  \mbox{$c\mid\!\!z\!\!\mid^{1/\delta_3}$}$
  with $\delta=\sum_i \delta_i$.
\end{references}
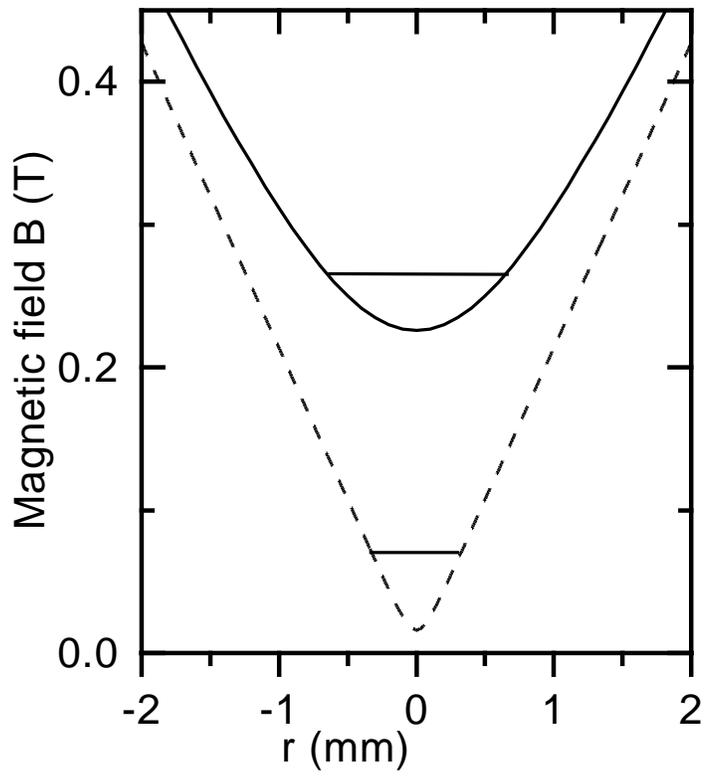
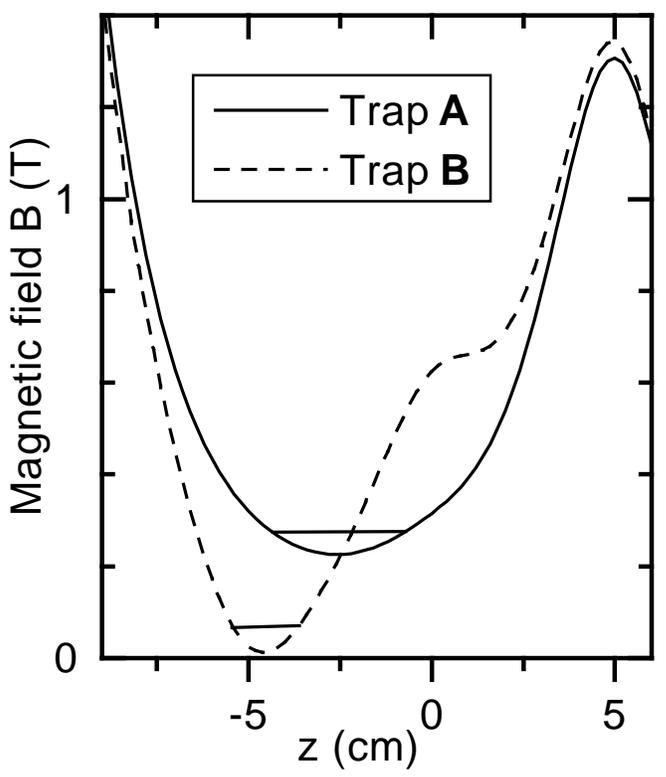
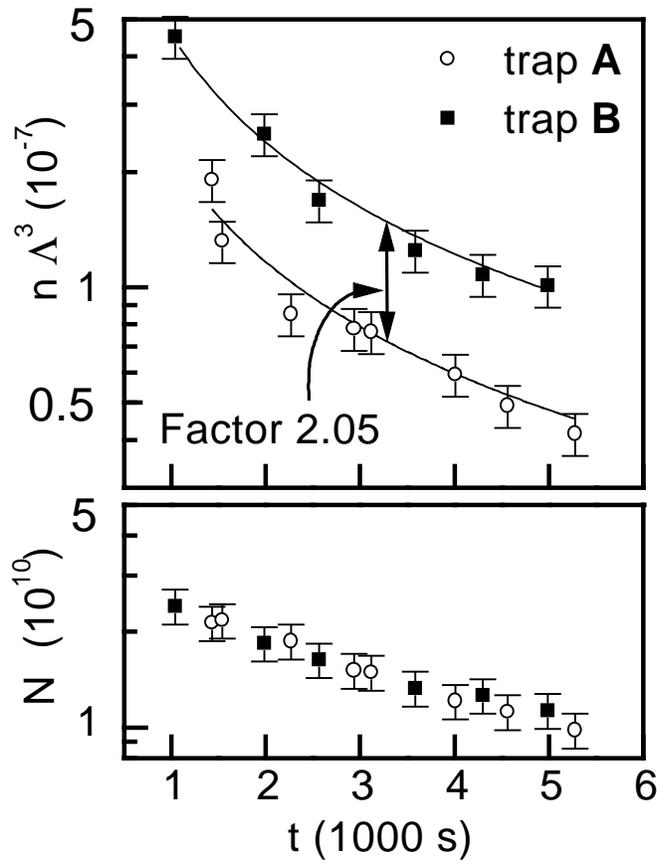
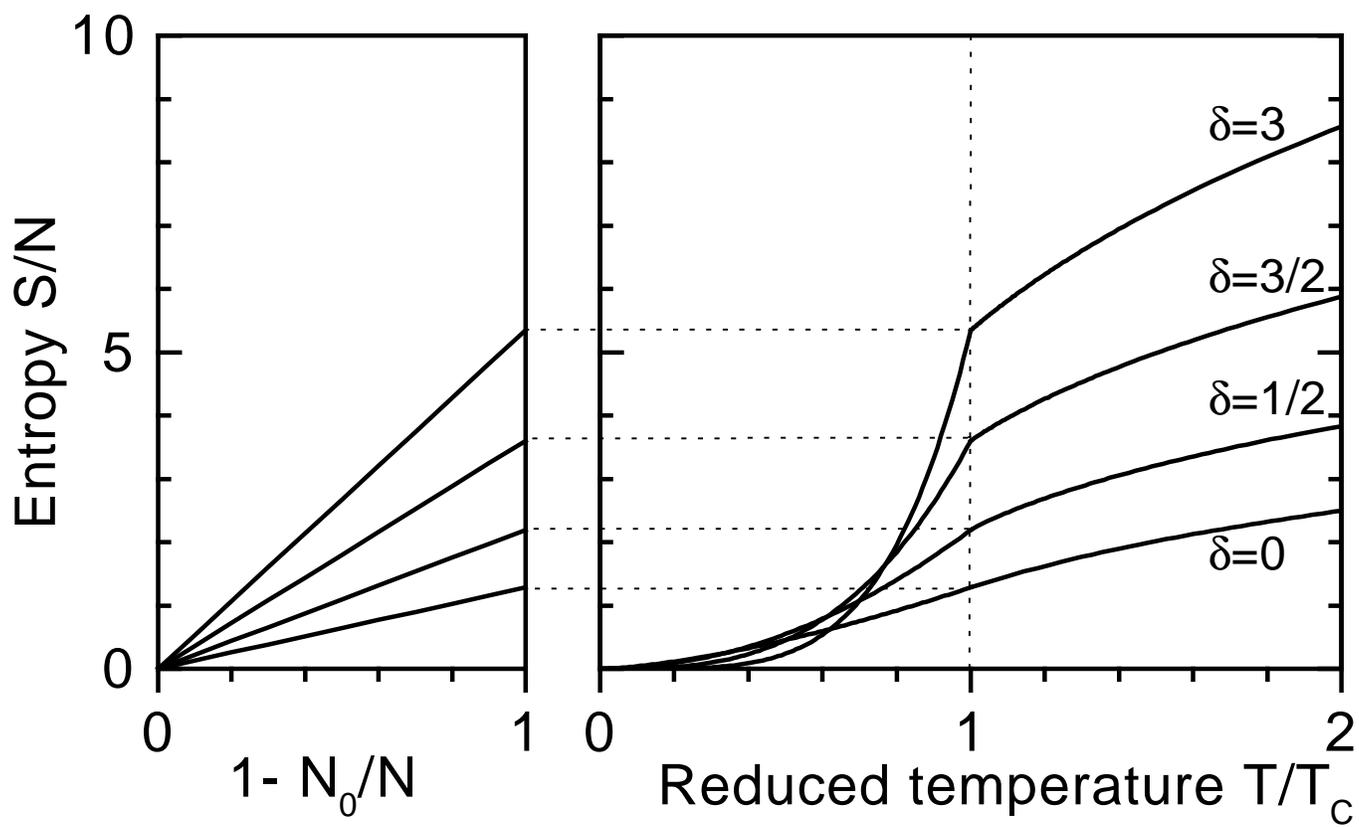
\end{document}